\documentclass[useAMS,usenatbib,usegraphicx]{emulateapj}
\usepackage{times}

\slugcomment{Astrophysical Journal Letters, in press (Accepted 2011 Aug 25).}

\shorttitle{Variable mid-IR jet emission in GX 339--4}
\shortauthors{P. Gandhi et al.}

\def\swift{{\sl Swift}}

\def\rxte{{\sl RXTE}}

\def\spitzer{{\em Spitzer}}

\def\p{$\pm$}
\def\ltsim{\mathrel{\hbox{\rlap{\hbox{\lower4pt\hbox{$\sim$}}}\hbox{$<$}}}}
\def\gtsim{\mathrel{\hbox{\rlap{\hbox{\lower4pt\hbox{$\sim$}}}\hbox{$>$}}}}

\def\Msun{M$_{\odot}$}
\def\micron{$\mu$m}
\def\araa{ARA\&A}
\def\aap{A\&A}

\def\aaps{A\&AS}
\def\mnras{MNRAS}
\def\apj{ApJ}
\def\apjs{ApJS}
\def\aj{AJ}
\def\apjl{ApJL}

\def\nub{$\nu_{\rm b}$}

\def\gx339{GX~339--4}
\def\av{$A_{\rm V}$}

\begin{document}

\title{A variable mid-infrared synchrotron break associated with the compact jet in GX 339--4}

\author{P. Gandhi\altaffilmark{1}, A.W. Blain\altaffilmark{2}, D.M. Russell\altaffilmark{3}, P. Casella\altaffilmark{4}, J. Malzac\altaffilmark{5,6}, S. Corbel\altaffilmark{7}, P. D'Avanzo\altaffilmark{8}, F.W. Lewis\altaffilmark{9}, S. Markoff\altaffilmark{3}, M. Cadolle Bel\altaffilmark{10}, P. Goldoni\altaffilmark{11,12}, S. Wachter\altaffilmark{13}, D. Khangulyan\altaffilmark{1} and A. Mainzer\altaffilmark{14}}
\altaffiltext{1}{Institute of Space and Astronautical Science, Japan Aerospace Exploration Agency, 3-1-1 Yoshinodai, chuo-ku, Sagamihara, Kanagawa 252-5210, Japan}
\altaffiltext{2}{Department of Physics \& Astronomy, University of Leicester, University Road, Leicester, LE1 7RH, UK}
\altaffiltext{3}{Astronomical Institute `Anton Pannekoek', University of Amsterdam, P.O. Box 94249, 1090 GE Amsterdam, the Netherlands}
\altaffiltext{4}{School of Physics and Astronomy, University of Southampton, Southampton, SO17 1BJ, UK}
\altaffiltext{5}{Universit\'{e} de Toulouse, UPS-OMP, IRAP, Toulouse, France}
\altaffiltext{6}{CNRS, IRAP, 9 Av. colonel Roche, BP 44346, F-31028 Toulouse cedex 4, France}
\altaffiltext{7}{Universit\'{e} Paris 7 Denis Diderot and Service d'Astrophysique, UMR AIM, CEA Saclay, F-91191 Gif sur Yvette, France}
\altaffiltext{8}{INAF-Osservatorio Astronomico di Brera, via Bianchi 46, I-23807, Merate (Lc), Italy}
\altaffiltext{9}{Faulkes Telescope Project, Division of Earth, Space and Environment, University of Glamorgan, Pontypridd CF37 1DL, Wales, UK}
\altaffiltext{10}{INTEGRAL Science Operations Centre Science Operations Department, European Space Astronomy Centre, Post Office Box 78, E-28691, Villanuevade la Cañada, Madrid, Spain}
\altaffiltext{11}{Laboratoire Astroparticule et Cosmologie, 10 rue A. Domon et L. Duquet, 75205 Paris Cedex 13, France}
\altaffiltext{12}{DSM/IRFU/Service d'Astrophysique, CEA/Saclay, 91191, Gif-sur-Yvette, France}
\altaffiltext{13}{Infrared Processing and Analysis Center, California Institute of Technology, Pasadena, CA 91125, USA}
\altaffiltext{14}{Jet Propulsion Laboratory, California Institute of Technology, Pasadena, CA 91109, USA}

\label{firstpage}
\begin{abstract}
Many X-ray binaries remain undetected in the mid-infrared, a regime where emission from their compact jets is likely to dominate. Here, we report the detection of the black hole binary GX 339--4 with the Wide-field Infrared Survey Explorer (WISE) during a very bright, hard accretion state in 2010. Combined with a rich contemporaneous multiwavelength dataset, clear spectral curvature is found in the infrared, associated with the peak flux density expected from the compact jet. An optically-thin slope of $\sim$--0.7 and a jet radiative power of $>$6$\times$10$^{35}$ erg s$^{-1}$ $(d/{\rm 8\ kpc)^2}$ are measured. A $\sim$24~h WISE light curve shows dramatic variations in mid-infrared spectral slope on timescales at least as short as the satellite orbital period $\sim$95 mins. There is also significant change during one pair of observations spaced by only 11~s. These variations imply that the spectral break associated with the transition from self-absorbed to optically-thin jet synchrotron radiation must be varying across the full wavelength range of $\sim$3--22~\micron\ that WISE is sensitive to, and more. Based on four-band simultaneous mid-infrared detections, the break is constrained to frequencies of $\approx$4.6$_{-2.0}^{+3.5}\times$10$^{13}$ Hz in at least two epochs of observation, consistent with a magnetic field $B$$\approx$1.5(\p0.8)$\times$10$^4$ G assuming a single-zone synchrotron emission region. The observed variability implies that either $B$, or the size of the acceleration zone above the jet base, are being modulated by factors of $\sim$10 on relatively-short timescales.
\end{abstract}

\section{Introduction}

Despite decades of studies on accreting sources, observations of jets in X-ray binaries suffer from important limitations. Strong emission from accreting, stellar or surrounding material dominates fainter jet radiation over infrared--X-ray frequencies in most sources \citep[e.g. ][]{russell06}. The transient nature of activity in many binaries makes it difficult to catch them contemporaneously with multiple observatories necessary to probe broadband emission and isolate the jet. Moreover, there has so far been a lack of sensitive monitoring instruments in the infrared where the jet flux density is expected to peak \citep[e.g. ][]{markoff01}. 

According to the canonical model of emission from compact jets \citep{blandford79}, this peak occurs because conservation of particle and magnetic flux leads to each location in the jet being associated with its own photospheric frequency, which falls linearly with distance from the base. The sum of all components conspires to give a flat/inverted spectrum ($\alpha \geq 0$, where $F_{\nu}\propto \nu^{\alpha}$) up to a peak frequency associated with the optically thick-to-thin break (\nub), above which optically-thin synchrotron radiation ($\alpha < 0$) reveals information about the underlying particle distribution. Constraints on this break remain sparse due to the limitations mentioned above \citep[e.g. ][]{fender01, corbel02, gallo07, migliari07, migliari10, peercasella09, rahoui11}. 

In this Letter, we present the first mid-infrared (MIR) study of GX 339--4, a binary hosting a black hole with mass $\gtsim$6 \Msun\ \citep{hynes03_gx339}, and well-known for its transient, multiwavelength variability on a broad range of timescales \citep{motch82,dunn08,g09_rmsflux,casella10}. Its donor star is a late-type companion in a long ($\sim$1.7 day) orbit, much fainter than the primary when active \citep{shahbaz01, munoz-darias08}. GX 339--4 shows evidence of relativistic jets and has proven to be a key object for jet studies across the electromagnetic spectrum \citep{fender01,corbel03,markoff03,g08}. 

A MIR detection with \spitzer\ has been reported, but only in a single filter during a faint, steep-power-law state \citep{tomsick04}. We now present multiband simultaneous MIR detections during the source outburst of 2010, which allow important new constraints on \nub, and on changing conditions in the inner jet. A distance of 8 kpc and a black hole mass of 10 \Msun\ are taken as representative  \citep{shidatsu11}. Interstellar line-of-sight extinction of \av=3.25 is assumed (Lewis et al., in prep.), with dereddening uncertainties described in detail.

\section{Observations}

\subsection{WISE}
\label{sec:wiseobs}

In the course of an all-sky survey, the Wide-field Infrared Survey Explorer (WISE) satellite \citep[][hereafter W10]{wise} detected GX 339--4 on 2010 March 11 (MJD 55266). This fortuitously occurred during a bright, hard state phase before outburst peak in April \citep[e.g. ][]{belloni11_atel}. 

WISE has four arrays (bands W1--W4 centered on wavelengths of 3.4, 4.6, 12 and 22 \micron, respectively) which image the same 47\arcmin$\times$47\arcmin\ field-of-view simultaneously using three dichroic beam splitters. The frame exposure times are 7.7 s for  W1--2, and 8.8 s for W3--4, respectively. Consecutive frames are separated by 11 s as the satellite scans across the sky. Though not designed for timing studies, the surveying strategy was such that every source was scanned multiple times over a period of $\sim$24 hours, sampling variability on timescales as short as the satellite orbital period. There is a $\sim$10\%\ field overlap between adjacent scans, thus sources may be caught on occasion only 11 s apart. 

The WISE all-sky survey preliminary data release\footnote{http://wise2.ipac.caltech.edu/docs/release/prelim/} reports Vega source magnitudes using both profile-fitting and standard aperture-corrected photometry. Profile-fitting represents the most accurate flux estimation for point sources for single-frame measurements\footnote{http://wise2.ipac.caltech.edu/docs/release/prelim/expsup/sec4\_3c.htm}, because the pipeline can account for image artifacts and source confusion. Some bad pixels are present around GX 339--4 but the field is free of diffraction spikes and detector latency caused by saturated bright sources. The point-spread function (PSF) corresponds to a Gaussian of about 6\arcsec.1--6\arcsec.5 (W1--W3) and 12\arcsec (W4) respectively, imaged at 2.8\arcsec/pix (W1--W3) and 5.6\arcsec/pix (W4). The two nearest bright stars to GX 339--4 lie $\sim$9 and 14\arcsec\ away, with mean central-pixel count-rates of $\sim$0.1 and 1.5 times GX 339--4 in W1 (and a decreasing count-rate towards W4). The pipeline successfully deblends multicomponent emission and reports acceptable $\chi^2$ values for the photometric fits. 

WISE absolute calibration is referenced to a network of standard stars \citep{cohen99}, and to sources at the ecliptic poles that are observed on most orbits. GX 339--4 magnitudes are listed in Table~\ref{tab:wisemags} for all 13 observing epochs. These were converted to fluxes using standard zeropoints and color corrections (W10) assuming a flat power-law. To account for present pipeline and absolute calibration uncertainties, we include a systematic zero point uncertainty of 2.4, 2.8, 4.5 and 5.7\% in W1--W4, respectively \citep{jarrett11}. The color correction mainly affects the W3 band, and we also include a 2\%\ uncertainty for this (W10), given the range of MIR spectral slopes discussed later. Improved calibration and reduced systematic errors may be expected after the full WISE data release in 2012.

\subsection{Multiwavelength}

Follow-up data will be presented in a number of upcoming papers, including: 1) Cadolle Bel et al. (2011; hereafter CB11) who describe extensive X-ray, near-IR and optical coverage; 2) Lewis et al. (in prep.) presenting further optical and UV data; and, 3) Corbel et al. (in prep.) discussing the radio. Broadband monochromatic fluxes closest to the WISE observations are listed in Table~\ref{tab:sed}, and plotted in Fig.~\ref{fig:sed}. Where multiple observations were available, averages were used for the mean spectral energy distribution (SED), with the mean error computed as the standard deviation divided by the square-root of the number of exposures. 

Below, we present some details on the X-ray analysis, the radio observations, and the UV--to--MIR dereddening, which are specific to our work. 

\subsubsection{X-rays}

Public X-ray data nearly-simultaneous with WISE from the {\em Rossi X-ray Timing Explorer} (\rxte; \citealt{rxte}) were available for analysis. A short observation (obsid 95409-01-09-03) with goodtime $\approx$1360 s was conducted between WISE epochs 12 and 13. For data reduction, we followed standard procedures using HEASOFT v.6.10 and FTOOLS (similarly to \citealt{g10}; hereafter G10). 

We find an acceptable fit with a multicolor disk with inner temperature $kT_{\rm in}$=0.80$_{-0.25}^{+0.15}$ keV (90\%\ confidence intervals), an Fe K line with equivalent-width ranging over $\sim$70--450 eV, and coronal Comptonization with electron temperature $kT_e$=32$_{-3}^{+4}$ keV and optical depth $\tau$=1.87$_{-0.20}^{+0.17}$ \citep[modelled with {\sc comptt};][]{comptt}. The PCA and HEXTE fluxes are found to be $F_{3-20\ \rm keV}$=3.70$_{-0.20}^{+0.01}$$\times$ 10$^{-9}$ and $F_{20-200\ \rm keV}$=8.6$_{-0.5}^{+0.1}$$\times$ 10$^{-9}$ erg s$^{-1}$ cm$^{-2}$, respectively, giving a luminosity $L_{1-200\ \rm  keV}$=1.0$\times$10$^{38}$ erg s$^{-1}$ or $\approx$0.08$\times$$L_{\rm Eddington}$. The bright-state spectrum of GX~339--4 is known to be complex and the exact physical model is irrelevant for the work presented here. The derived luminosity is not overly sensitive to model parametrization. See CB11 for detailed X-ray evolution of the source over the outburst. 

For timing analysis, we extracted net light curves from the \rxte\ event mode data, with 8 s time bins in order to approximately match the WISE exposure times. \rxte\ data probe variability shorter than 1360 s. For longer timescales, we used \swift\ Burst Alert Telescope (BAT) 15--50 keV light curves provided by the Hard X-ray Transient Monitor team\footnote{http://swift.gsfc.nasa.gov/docs/swift/results/transients/weak/GX339-4}. \swift\ sampling times are non-uniform, with intervals ranging from $\sim$10 mins to $\gtsim$1~h.

\subsubsection{Radio}

The Australia Telescope Compact Array (ATCA) observed GX 339--4 frequently during outburst. Here, we have used the mean fluxes observed on the two closest dates straddling the WISE observation: MJD 55262.91 and 55269.80, with simultaneous observations in the 5.5 and 9 GHz bands. Variations of $\approx$25 and 40\% are present between the bands. Note that the source does not display extreme radio variability on timescales of days \citep{corbel00}. Full details will be in Corbel et al. (in prep.).

\subsubsection{Dereddening}

The UV--to--near-IR data were dereddened assuming a standard interstellar extinction law with $R_{\rm V}$=3.1 \citep{cardelli89}. Extinction beyond 3~\micron, though certainly lower than in the near-IR, remains ill-constrained. We use the mean of three recent extinction laws for the WISE data, with their dispersion serving as a propagated systematic uncertainty on the corrected fluxes: 1) the $R_{\rm V}$=5.5 grain model of \citet{draine03} which matches the Galactic center extinction of \citet{lutz96} below 8~\micron; 2) the Galactic center extinction curve described by \citet{chiar06}; and 3) that presented by \citet{flaherty07}. The latter two provide updates on MIR Silicate absorption strengths. Finally, we also include a systematic \av\ uncertainty of 0.5 mags (G10).

\section{Results}

\subsection{Broad-band spectrum}

GX 339--4 is detected with high significance in all four WISE bands, with average signal:noise of $\sim$40--50 in W1--W3 and 22 in W4 across epochs. With mean observed fluxes of $\gtsim$60 mJy in W3 and W4, these detections are amongst the brightest for transient binary systems in the MIR \citep{wachter08, vrtilek08}, especially when considering the isolated jet component, as we discuss below.

The SED in Fig.~\ref{fig:sed} shows two peaks. A rising spectrum in the optical/ultraviolet (OUV) may originate either via pre-shock jet synchrotron or accretion disk (re)radiation (cf. G10). Dominant dereddening uncertainty prevents detailed inferences here, but we note that the optical--to--X-ray flux ratio is relatively low at $\lambda F_\lambda$(5500~\AA)/$F_{2-10}$$\sim$0.08 (with a factor of two uncertainty), so disk X-ray irradiation could provide the requisite power. 

A second peak is in the MIR, associated with broad spectral curvature declining towards the near-IR and to the radio. We discuss below that this curvature is likely to be emission from the compact jet, with the WISE W1--2 bands preferentially probing optically-thin radiation. Thus fitting a two-power-law model (one for the optically-thin jet and one for the OUV) over the W2--UV frequency range yields a mean optically-thin spectral slope $\alpha_{\rm thin}$=--0.73\p0.24, and a largely unconstrained OUV slope $\alpha_{\rm OUV}$=$+$1.3\p0.9. Fitting a single optically-thick power-law to the radio--W4 regime yields $\alpha_{\rm thick}$=$+$0.29\p0.02, though the average SED indicates that a single power-law may be an over-simplification. Errors include systematic uncertainties. 

$\alpha_{\rm thin}$ measures the distribution function of jet plasma energies, implying a power-law slope $p$=1--2$\alpha_{\rm thin}$=2.5\p0.5 \citep{rybickilightman}. Integrating over the mean radio--near-IR (2 \micron) SED using piece-wise power-law interpolation gives a lower limit to the jet radiative power ($L_{\rm rad}$) of 6$\times$10$^{35}$ erg s$^{-1}$, assuming that corrections due to relativistic beaming are of order unity \citep{casella10}. If the X-ray power ($L_{1-200\ \rm keV}$) is close to the {\em total} jet power (\citealt{koerding06}), we may place a lower limit on the jet radiative efficiency, $\eta$=$L_{\rm rad}$/$L_{\rm jet}$$\gtsim$1\%.

\subsection{Variability}

Simultaneous four-band WISE light curves in Figure~\ref{fig:lcs} show striking variability typically sampled at (multiples of) the WISE orbital period of ~95 mins. The shortest sampling interval was 11 s between epochs 3 and 4, when WISE caught GX 339--4 on two consecutive scans. We extracted light curves of randomly-selected field stars for comparison. For each band, two field stars with brightness similar to GX 339--4 within 5$'$ were selected. Variations in the target light curves are clearly much stronger than the field star variations and also display inter-band correlations, implying that they are not related to any instrumental or observational artifact. 

We measured the fractional variability amplitudes ($F_{\rm var}$; \citealt{vaughan03}) of the MIR light curves and found them to increase with wavelength, with $F_{\rm var}^{\rm MIR}$$\sim$0.25\p0.03 for W1--2 and $\sim$0.32\p0.06 for W3--4, respectively (errors allow for WISE pipeline systematics as in \S~\ref{sec:wiseobs}). The peak-to-peak flux ratios also increase from 2.3--3.3 in W1--4. In X-rays, the \swift\ data show low mean variability over the timespan of WISE observations, with $F_{\rm var}$(\swift)=0.17\p0.02. This value may suffer from large systematics because BAT employs a coded-mask with a wide field-of-view. \rxte\ is not affected by such complications, and we measure $F_{\rm var}$(PCA)=0.21\p0.01 and $F_{\rm var}$(HEXTE)=0.20\p0.01 for the full energy ranges of the instruments, and over the range of timescales of 8--1360~s.

\section{Discussion}
\label{sec:discussion}

Both the timing and spectral properties strongly favor a jet origin for the mid-infrared emission. The mean optically-thick and thin slopes are fairly typical for binaries in the hard state when the jet is known to be strong. From closely-simultaneous X-ray and MIR light curves, we find $F_{\rm var}^{\rm X-ray}$$<$$F_{\rm var}^{\rm MIR}$ when comparing similar timescales. This argues against strong MIR variations originating via reprocessing, because the secondary (driven) variability cannot exceed the primary ones in strength. X-ray variations can also be ruled out as the driver of IR variations in the outer disk because the IR spectral slopes are inconsistent with thermal reprocessing. Dusty circumbinary disks found in some other sources \citep{rahoui10} are unlikely here because of the strong variability (so dust grains should be destroyed by X-rays), and the MIR spectral slopes do not match thermal dust emission.

\begin{figure*}
  \begin{center}
  \includegraphics[width=12.5cm]{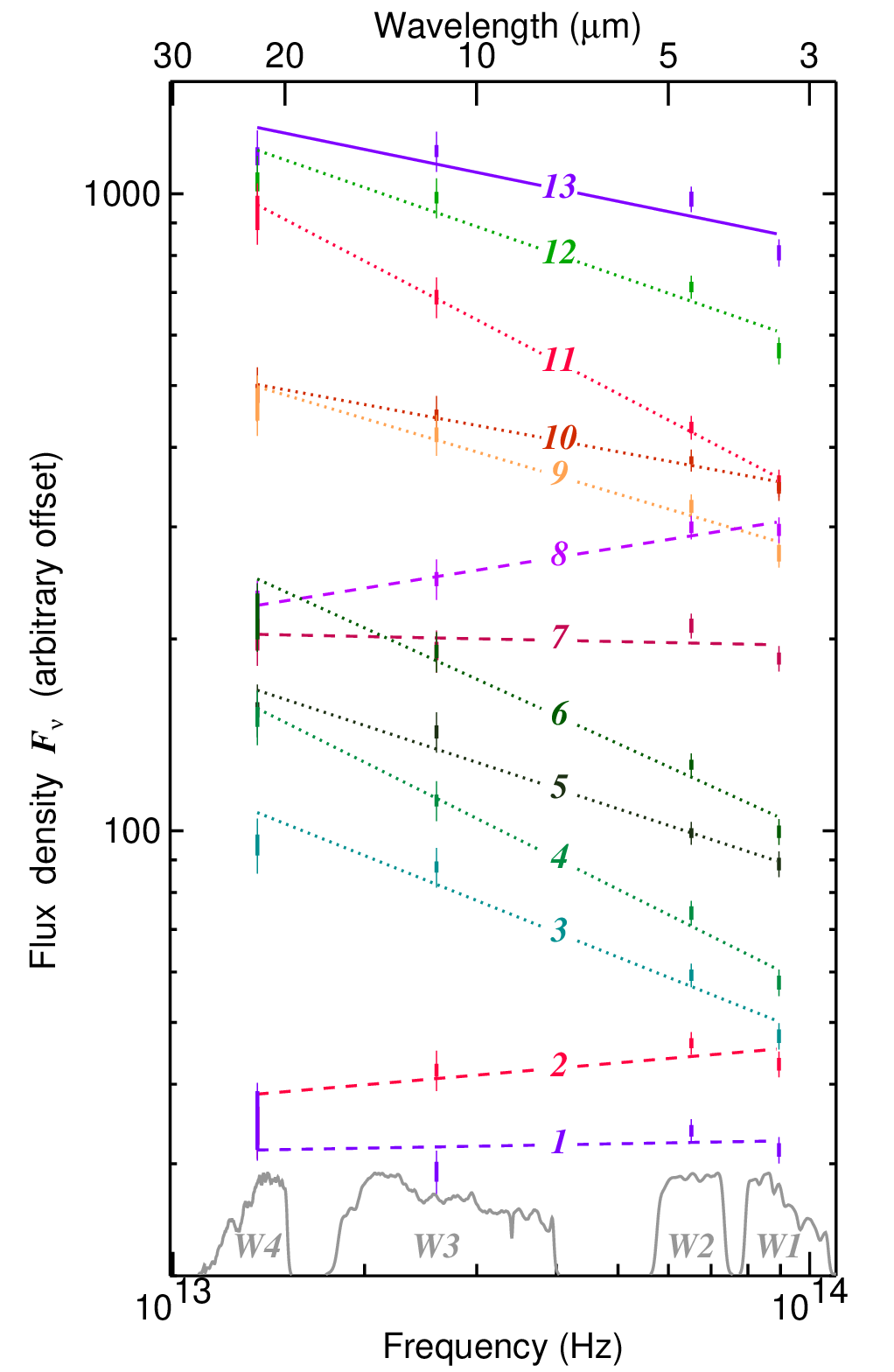}
  \caption{Evolution of the WISE spectra across epochs (labeled) with single-power-law fits overplotted. Epochs 1--13 have multiplicative offsets of 0.8, 0.6, 0.8, 1.3, 1.4, 2.5, 4, 6, 5, 5, 10, 9 and 10, respectively, for clarity. Thick error bars are statistical uncertainties, and thin ones include systematic uncertainties. Response profiles of the WISE bands are plotted at the bottom, arbitrarily normalized. 
\label{fig:epochseds}}
\end{center}
\end{figure*}

The strong variability implies dramatic changes in spectral slope across the WISE bands in the individual observation epochs. These are displayed in Fig.~\ref{fig:epochseds}. Parametrizing the spectra with single power-laws results in slopes of $\alpha_{\rm WISE}$$\sim$--0.5 to +0.2 (Table~\ref{tab:wisemags}). In several epochs, additional spectral curvature appears to be present within the WISE bands. In particular, in epochs 12 and 13, W1 and W4 fluxes lie below the fit (while W2--3 lie above), returning large single-power-law null-hypothesis rejection probabilities ($P$). From W2 to W1, the slope is negative in most epochs ($\overline{\alpha_{{}_{\rm W1}}^{{}_{\rm W2}}}$=--0.48\p0.07), meaning that these bands probe mostly optically-thin emission. Thus carrying out a simple two-power-law fit to describe the W1 \& W2 band-pair and the W3 \& W4 pair separately, the spectral break associated with this curvature is constrained to be at log(\nub; Hz)=13.65\p0.24 and 13.68\p0.26 in epochs 12 and 13, respectively. 

The position of \nub\ depends strongly on the magnetic field strength ($B$) at the acceleration zone in the compact jet, and is comparatively insensitive to luminosity, geometry and energy partition factors. Generalizing the homogeneous cylindrical single-zone emission model of \citet{chaty11}, we find expressions for $B$ and the zone radius $R$ as follows.

\begin{eqnarray}
  \nonumber
  B & = & 1.5\times 10^4\ \  \xi^{-4z}\ h^{2z}\ d_8^{-4z} \\
    &   & \times\ (F_{\nu_{\rm b}}/F_{70})^{-2z}\ (\nu_{\rm b}/4.5\times10^{13})\ {\rm G},
\end{eqnarray}
\begin{eqnarray}
  \nonumber
  R & = & 2.2\times 10^9\ \  \xi^{-z}\ h^{-z(p+6)}\ d_8^{2z(p+6)}\\
    &   & (F_{\nu_{\rm b}}/F_{70})^{z(p+6)}\ (\nu_{\rm b}/4.5\times10^{13})^{-1}\ {\rm cm},
\end{eqnarray}

\noindent
where $z$=(2$p$+13)$^{-1}$, $\xi$ is the equipartition fraction of particles to magnetic energy density, $h$ is the scale-height of the optically-thin region, $d_8$ is the distance in units of 8 kpc, and $F_{70}$=70 mJy. Since $B$ depends only weakly on $h$ and $\xi$ (we fix both at 1), we have a robust estimate of $B$ when \nub\ can be measured. Both epochs 12 and 13 imply $B$$\approx$1.5(\p0.8)$\times$10$^4$ G and $R$$\approx$2.5(\p1.5)$\times$10$^9$ cm, with errors determined from Monte Carlo sampling of all parameters (including also systematic distance uncertainty of 2 kpc).

In addition, the strong MIR slope changes between optically-thick and thin imply that \nub\ is varying over a range at least as wide as that which WISE is sensitive to. It is likely that uncertainties due to present WISE photometric calibration as well as the ill-constrained dereddening prevent accurate localization of \nub\ in some other epochs. Setting the WISE systematic errors to zero and repeating the single-power-law fits returns $P$$>$90\% values in additional epochs 1, 2, 3, 6 and 7, consistent with \nub\ falling within the WISE bandpass in these. 

In order to understand the extent of changing jet physical conditions, we compare two extreme WISE epochs: 1) epoch 4, with very bright flux in W4 and a steep slope implying optically-thin emission over the whole WISE range; 2) epoch 8, when the source was faintest in W4 with an inverted slope implying that \nub\ has shifted to the high frequency end. We assume 1) $\nu_{\rm b}$(epoch 4)=1.3$\times$10$^{13}$ Hz and 2) $\nu_{\rm b}$(8)=9$\times$10$^{13}$ Hz, corresponding to the extreme WISE frequencies and representing likely upper and lower limits to the true \nub\ values, respectively. The dereddened fluxes are $F_{\nu_{\rm b}}$(4)$\approx$115 mJy and $F_{\nu_{\rm b}}$(8)$\approx$50 mJy. Comparing $B$ and $R$, we find

\begin{equation}
B(4)/B(8) = [F_{\nu_{\rm b}}(4)/F_{\nu_{\rm b}}(8)]^{-2z} [\nu_{\rm b}(4)/\nu_{\rm b}(8)] = 0.13
\end{equation}
\begin{equation}
R(4)/R(8) = [F_{\nu_{\rm b}}(4)/F_{\nu_{\rm b}}(8)]^{z(p+6)} [\nu_{\rm b}(4)/\nu_{\rm b}(8)]^{-1} = 10.4
\end{equation}

\begin{figure*}
  \begin{center}
  \includegraphics[width=8.0cm]{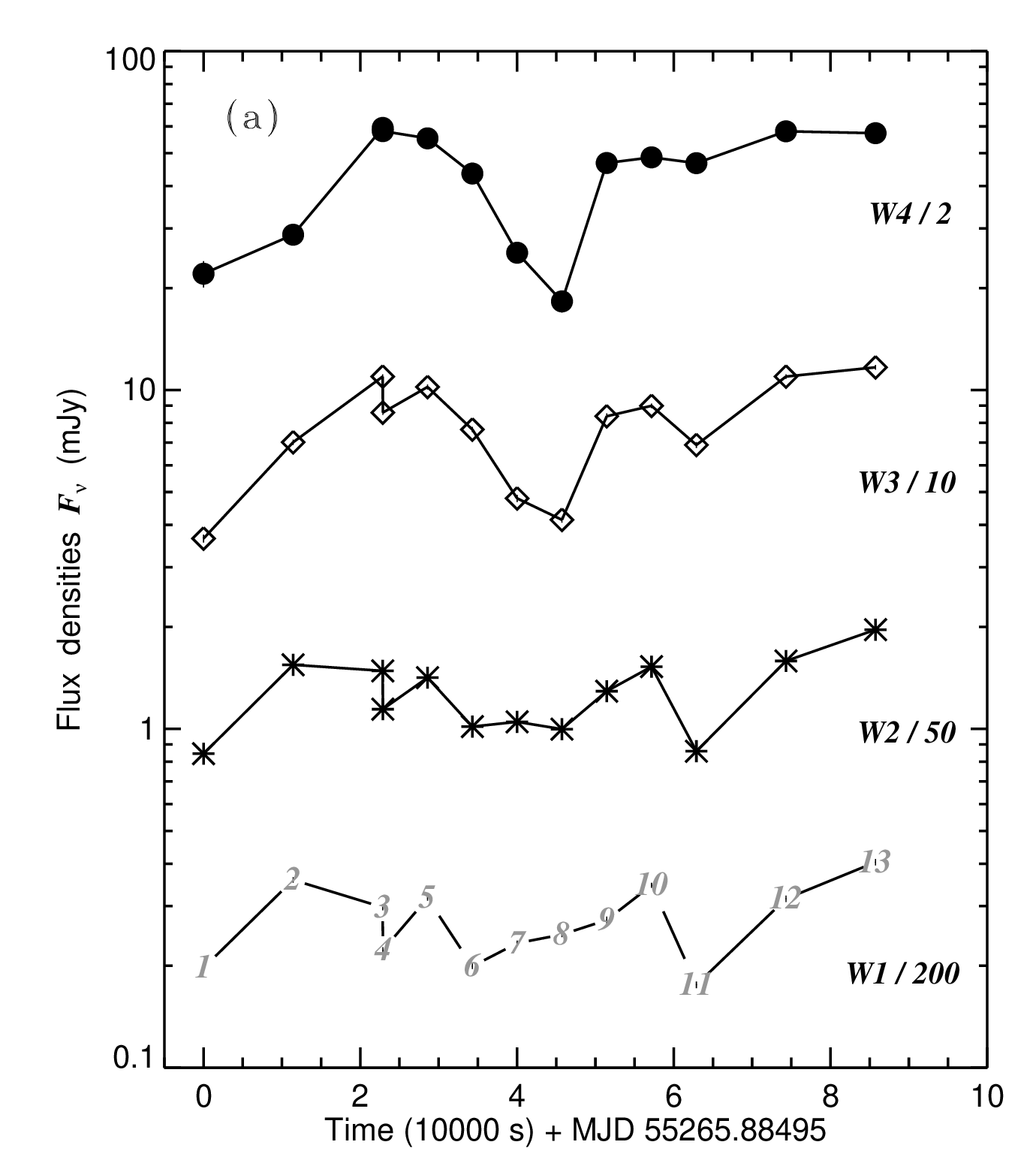}
  \includegraphics[width=8.0cm]{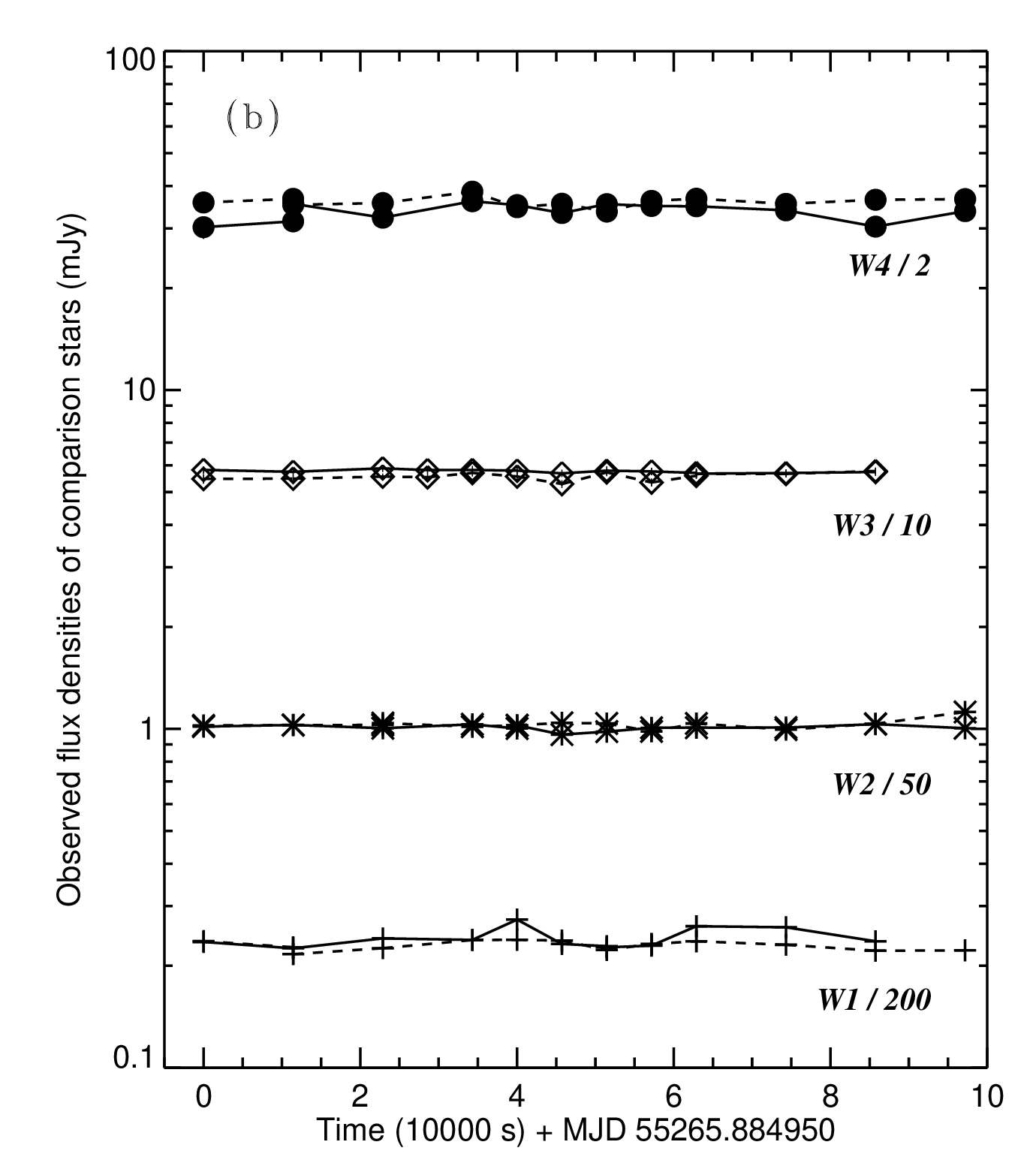}
  \includegraphics[width=8.0cm]{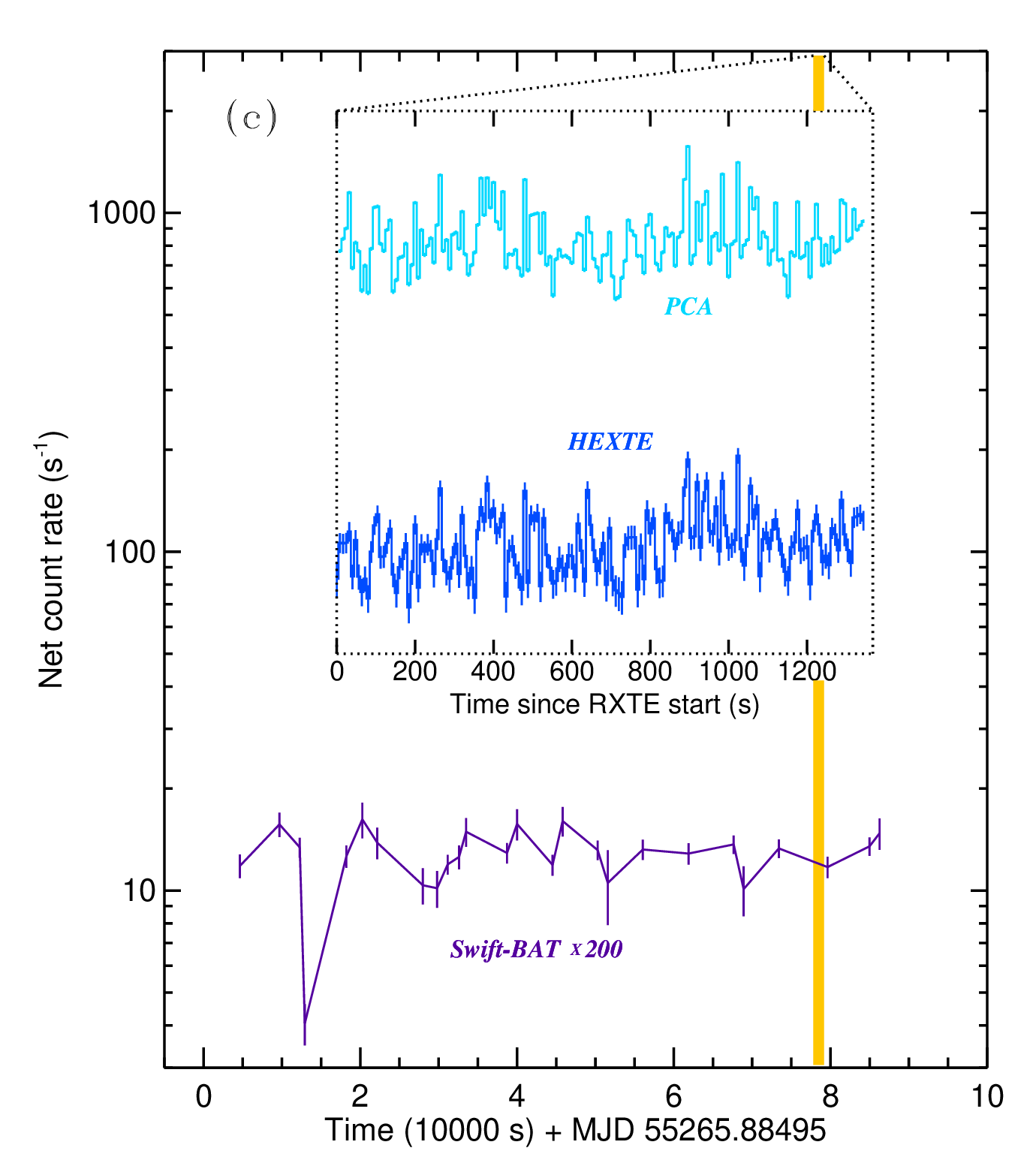}
  \caption{{\em (a)} WISE flux light curves of GX 339--4 showing the 13 epochs of measurement. Statistical error bars are typically smaller than the plot symbols. Band light curves have been offset by the labeled factors for clarity. {\em (b)} Light curves for randomly-selected comparison field stars (two stars per band are shown as the solid and dashed lines, respectively). {\em (c)} X-ray light curves. The 15--50 keV \swift/BAT hard X-ray light curve covers a full day. The \rxte\ observing period is denoted in yellow, with the inset showing the PCA and HEXTE light curves in detail. The y-axis has an identical relative scaling in all panels.
    \label{fig:lcs}}
  \end{center}
\end{figure*}

\noindent
Modulations by factors of $\sim$10 in the field strength or in acceleration zone size are required. The standard scaling of the $B$ field along the jet is $B$$\sim$$R$$^{-1}$ \citep{blandford79} which, if true, is consistent with the variability being dominated by changes in location of the acceleration zone along the jet. These modulations may ultimately originate from stochastic instabilities within the accretion flow, or as a result of changing dissipative shock conditions in the inner jet \citep[e.g., ][]{falcke95, meier01, polko10}. Multiwavelength monitoring coordinated with the mid-infrared opens up the possibility for studying these processes in the future. The characteristic timescales of change are $\sim$hours or less (Fig.~\ref{fig:lcs}), though variability between epochs 3 and 4 means that some changes are faster than $\sim$11 s.

\begin{figure*}
  \begin{center}
  \includegraphics[width=13cm,angle=90]{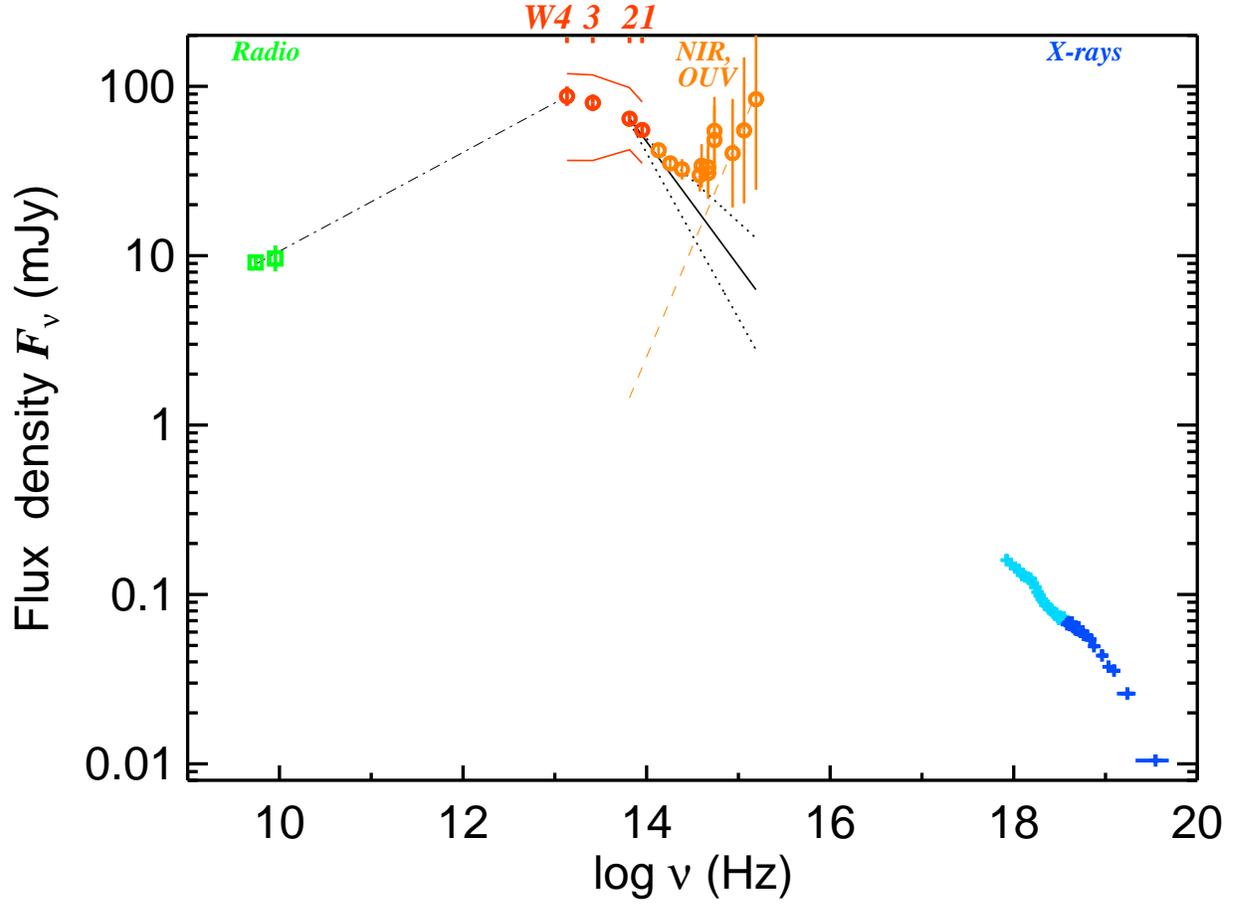}
  \caption{Average dereddened SED of GX 339--4. WISE data points are in red; red curves represent the envelope of extreme variations over 13 WISE epochs. The near-infrared and optical/ultraviolet (OUV) points are plotted in orange, radio in green. The optically-thin jet power-law is shown as the solid black line, and 1$\sigma$ fit uncertainties on this slope by the dotted lines. The dashed and dot-dashed lines represent the OUV and optically-thick power-laws, respectively. Detailed modeling of the origin of X-rays (blue; unfolded best-fit model) is beyond the scope of this Letter.
    \label{fig:sed}}
  \end{center}
\end{figure*}

Simultaneous multiband observations over a broad frequency range with WISE have proved crucial for revealing complex and rapid changes occurring near the jet base. Based on the average data in Fig.~\ref{fig:sed} alone, it would be impossible to accurately localize \nub\ and to infer changes in its position. Whereas \nub\ has been found to lie at similar MIR frequencies of (1--4)$\times$10$^{13}$ Hz in 4U 0614+091 and Cyg X-1, observational limitations prevented simultaneous observations straddling the break in the former \citep{migliari10}, and the MIR is dominated by non-jet emission in the latter \citep{rahoui11}. We have cleanly probed jet-dominant MIR emission simultaneously in multiple bands and shown that inferences from long-term averages are not representative of the underlying behavior in GX 339--4, at least. Thus, time dependence must be incorporated as an integral parameter when modeling compact jets. 

Scale-invariant jet models imply that similar break frequency variability could be present in active galactic nuclei, where \nub\ is expected at mm frequencies \citep{kellermann69, heinz03}. Long-term mm monitoring may have already found signatures for this \citep{trippe11}.

\acknowledgements
WISE is a joint project of the University of California, Los Angeles, and the Jet Propulsion Laboratory (JPL)/California Institute of Technology (Caltech), funded by the National Aeronautics and Space Administration (NASA). Data for individual scans are from NEOWISE \citep[e.g. ][]{neowise}, a project of JPL/Caltech funded by NASA's Planetary Science Division. We thank the referee for a prompt report. Individual support acknowledgments are as follows. PGa: JAXA International Top Young Fellowship; DMR and SM: Netherlands Organisation for Scientific Research Veni and Vidi Fellowship, respectively; JM: GDR PCHE (France); PC: EU Marie Curie Intra-European Fellowship \#2009-237722. \swift/BAT transient monitor results and \rxte\ HEASARC archive data are used herein. The Faulkes Telescope South (FTS) is maintained and operated by Las Cumbres Observatory Global Telescope Network. We thank Rosa Doran of NUCLIO, Portugal for FTS observations as part of the EU-Hands on Universe initiative for school scientific education.

\clearpage

\begin{deluxetable}{lccccccr}
\tabletypesize{\scriptsize}
\tablecaption{GX 339--4 WISE photometry \label{tab:wisemags}}
\tablewidth{0pt}
\tablecolumns{8} 
\tablehead{
  \colhead{Epoch}       &
  \colhead{MJD}         &
  \colhead{W1$^\S$}         &
  \colhead{W2$^\S$}         &
  \colhead{W3$^\S$}         &
  \colhead{W4$^\S$}         &  
  \colhead{$\alpha_{\rm WISE}^{*}$} &
  \colhead{$P$$^{\#}$}        \\
  \colhead{}&
  \colhead{}&
  \colhead{mag~~~~~(mJy)}         &
  \colhead{mag~~~~~(mJy)}         &
  \colhead{mag~~~~~(mJy)}         &
  \colhead{mag~~~~~(mJy)}         &
  \colhead{}&
  \colhead{\%}             }
\startdata
1   & 55265.8849  & 9.96\p0.03 (39.4\p2) & 9.19\p0.02 (42.3\p2) & 7.57\p0.04 (36.5\p3) & 5.86\p0.11 (44.1\p7)  & +0.02  & 82 \\
2   & 55266.0173  & 9.31\p0.02 (71.7\p3)  & 8.54\p0.02 (77.3\p3) & 6.86\p0.03 (70.1\p5) & 5.57\p0.07 (57.5\p7)  & +0.09  & 86 \\
3   & 55266.1496  & 9.51\p0.03 (59.5\p3)  & 8.58\p0.02 (74.2\p3) & 6.37\p0.02 (109.6\p8) & 4.79\p0.04 (118.8\p12) & --0.40 & 88 \\
4   & 55266.1497  & 9.83\p0.03 (44.4\p2)  & 8.87\p0.03 (57.2\p3) & 6.64\p0.02 (85.8\p6) & 4.81\p0.04 (116.2\p11) & --0.50 & 66 \\
5   & 55266.2158  & 9.45\p0.02 (63.3\p3)  & 8.63\p0.02 (70.8\p3) & 6.45\p0.03 (102.1\p7) & 4.86\p0.03 (110.6\p11) & --0.33 & 48 \\
6   & 55266.2820  & 9.95\p0.02 (39.9\p2)  & 8.99\p0.02 (50.8\p2) & 6.76\p0.03 (76.5\p6) & 5.12\p0.09 (87.1\p11) & --0.46 & 85 \\
7   & 55266.3481  & 9.78\p0.02 (46.6\p2)  & 8.96\p0.03 (52.5\p2) & 7.27\p0.04 (47.9\p4) & 5.71\p0.06 (50.9\p 5) & --0.02 & 82 \\
8   & 55266.4143  & 9.72\p0.03 (49.4\p2)  & 9.01\p0.02 (49.9\p2) & 7.43\p0.03 (41.4\p3) & 6.07\p0.09 (36.5\p 5) & +0.16  & 40 \\
9   & 55266.4804  & 9.61\p0.03 (54.6\p3)  & 8.73\p0.03 (64.6\p3) & 6.67\p0.03 (83.7\p6) & 5.05\p0.06 (93.6\p10) & --0.30 & 56 \\
10  & 55266.5466  & 9.35\p0.02 (69.1\p3)  & 8.55\p0.01 (76.3\p3) & 6.59\p0.02 (89.9\p6) & 5.00\p0.04 (97.3\p10) & --0.19 & 25 \\
11  & 55266.6128  & 10.09\p0.03 (35.1\p2) & 9.18\p0.02 (43.0\p2) & 6.88\p0.03 (68.9\p5) & 5.05\p0.07 (93.5\p10) & --0.52 & 17 \\
12  & 55266.7451  & 9.45\p0.03 (63.1\p3)  & 8.51\p0.02 (79.3\p3) & 6.37\p0.02 (109.6\p8) & 4.81\p0.04 (116.1\p11) & --0.35 & 93 \\
13  & 55266.8774  & 9.18\p0.03 (80.9\p4)  & 8.28\p0.03 (98.1\p5) & 6.30\p0.02 (116.7\p8) & 4.83\p0.04 (114.6\p11) & --0.21 & 93 \\
\enddata
\tablenotetext{\S}{Mags and statistical errors reported in the WISE preliminary data release, based upon profile-fitting. Numbers in brackets are dereddened and color-corrected fluxes, with errors including systematics.} 
\tablenotetext{*}{Spectral slope from single-power-law fit to each flux-calibrated WISE epoch. Uncertainties are $\approx$0.04--0.05 (1$\sigma$).}
\tablenotetext{\#}{Null hypothesis probability at which a single-power-law fit may be rejected, based upon $\chi^2$ statistics with two degrees-of-freedom.}
\end{deluxetable}

\begin{deluxetable}{lcccr}
\tabletypesize{\scriptsize}
\setlength{\tabcolsep}{0.04in}
\tablecaption{Quasi-simultaneous dereddened ultraviolet to radio mean fluxes \label{tab:sed}}
\tablewidth{0pt}
\tablehead{
\colhead{Band (Telescope or Instrument)$^a$}       &
\colhead{log Frequency}         &
\colhead{Mean MJD$^b$}         &
\colhead{Flux density$^c$}         &
\colhead{Reference}         \\
\colhead{}        &
\colhead{Hz}        &
\colhead{}        &
\colhead{mJy}        
}
\startdata
UVW2 (UVOT) & 15.19 & 55261.05 & 83.8(\p6.9)$_{-59}^{+203}$ & CB11; Lewis et al. (in prep.)\\
UVW1 (UVOT) & 15.06 & 55259.97 & 54.6(\p3.0)$_{-35}^{+93}$ & CB11; Lewis et al. (in prep.)\\
U (UVOT)    & 14.94 & 55260.38 & 40.3(\p0.7)$_{-21}^{+44}$ & CB11; Lewis et al. (in prep.)\\
V (FTS)     & 14.74 & 55273.72 & 54.5(\p1.8)$_{-20}^{+32}$ & CB11; Lewis et al. (in prep.) \\
R (FTS)     & 14.67 & 55273.72 & 33.2(\p2.1)$_{-10}^{+14}$ & CB11; Lewis et al. (in prep.)\\
i' (FTS)    & 14.60 & 55273.72 & 33.9(\p0.3)$_{-9}^{+12}$ & CB11; Lewis et al. (in prep.)\\
V (REM)     & 14.74 & 55261.30 & 48.2(\p1.1)$_{-18}^{+28}$ & CB11\\
R (REM)     & 14.67 & 55261.30 & 30.7(\p0.7)$_{-9}^{+13}$ & CB11\\
I (REM)     & 14.58 & 55261.30 & 29.9(\p0.3)$_{-6}^{+7}$ & CB11\\
J (REM)     & 14.39 & 55261.30 & 32.4(\p0.8)$_{-4}^{+5}$ & CB11\\
H (REM)     & 14.26 & 55261.30 & 35.1(\p0.3)$_{-3}^{+3}$ & CB11\\
K (REM)     & 14.13 & 55261.30 & 41.9(\p0.7)$_{-3}^{+3}$ & CB11\\
W1 (WISE)   & 13.95 & 55266.363 & 55.2(\p3.9)$_{-4}^{+5}$ & This work\\
W2 (WISE)   & 13.81 & 55266.363 & 64.3(\p4.6)$_{-5}^{+5}$ & This work\\
W3 (WISE)   & 13.41 & 55266.363 & 79.9(\p7.3)$_{-9}^{+10}$ & This work\\
W4 (WISE)   & 13.13 & 55266.363 & 87.4(\p8.3)$_{-11}^{+12}$ & This work\\
9 GHz (ATCA) & 9.95  & 55266.4   & 9.7(\p0.1)\p1.7 & Corbel et al. (in prep.)\\
5.5 GHz (ATCA) & 9.74  & 55266.4   & 9.1(\p0.1)\p1.0 & Corbel et al. (in prep.)\\
\enddata
\tablenotetext{a}{UVOT=\swift\ UltraViolet and Optical Telescope; FTS=Faulkes Telescope South; REM=Rapid Eye Mount telescope} 
\tablenotetext{b}{Closest available data to the WISE observation.}
\tablenotetext{c}{Dereddened fluxes. Statistical error-bars are in brackets; the following numbers are the total errors including systematics.}
\end{deluxetable}

\label{lastpage}

\begin{thebibliography}{50}
\expandafter\ifx\csname natexlab\endcsname\relax\def\natexlab#1{#1}\fi

\bibitem[{{Belloni} {et~al.}(2010){Belloni}, {Motta}, \&
  {Mu{\~n}oz-Darias}}]{belloni11_atel}
{Belloni}, T., {Motta}, S., \& {Mu{\~n}oz-Darias}, T. 2010, The Astronomer's
  Telegram, 2577, 1

\bibitem[{{Blandford} \& {K\"{o}nigl}(1979)}]{blandford79}
{Blandford}, R.~D., \& {Konigl}, A. 1979, \apj, 232, 34

\bibitem[{{Bradt} {et~al.}(1993){Bradt}, {Rothschild}, \& {Swank}}]{rxte}
{Bradt}, H.~V., {Rothschild}, R.~E., \& {Swank}, J.~H. 1993, \aaps, 97, 355

\bibitem[{{Cadolle Bel} {et~al.}(2011){Cadolle Bel}, {Rodriguez}, {D'Avanzo}, {Russell}, {Tomsick}, {Corbel}, {Lewis}, {Rahoui}, {Buxton}, {Goldoni}, \& {Kuulkers}}]{cadollebel11}
{Cadolle Bel}, M., {Rodriguez}, J., {D'Avanzo}, P., {Russell}, D.~M., {Tomsick}, J., {Corbel}, S., {Lewis}, F., {Rahoui}, F., {Buxton}, M., {Goldoni}, P. \& {Kuulkers} E. 2011, A\&A, submitted (CB11)

\bibitem[{{Cardelli} {et~al.}(1989){Cardelli}, {Clayton}, \&
  {Mathis}}]{cardelli89}
{Cardelli}, J.~A., {Clayton}, G.~C., \& {Mathis}, J.~S. 1989, \apj, 345, 245

\bibitem[{{Casella} {et~al.}(2010){Casella}, {Maccarone}, {O'Brien}, {Fender},
  {Russell}, {van der Klis}, {Pe'Er}, {Maitra}, {Altamirano}, {Belloni},
  {Kanbach}, {Klein-Wolt}, {Mason}, {Soleri}, {Stefanescu}, {Wiersema}, \&
  {Wijnands}}]{casella10}
{Casella}, P., {Maccarone}, T.~J., {O'Brien}, K., {Fender}, R.~P., {Russell},
  D.~M., {van der Klis}, M., {Pe'Er}, A., {Maitra}, D., {Altamirano}, D.,
  {Belloni}, T., {Kanbach}, G., {Klein-Wolt}, M., {Mason}, E., {Soleri}, P.,
  {Stefanescu}, A., {Wiersema}, K., \& {Wijnands}, R. 2010, \mnras, 404, L21

\bibitem[{{Chaty} {et~al.}(2011){Chaty}, {Dubus}, \& {Raichoor}}]{chaty11}
{Chaty}, S., {Dubus}, G., \& {Raichoor}, A. 2011, \aap, 529, A3+

\bibitem[{{Chiar} \& {Tielens}(2006)}]{chiar06}
{Chiar}, J.~E., \& {Tielens}, A.~G.~G.~M. 2006, \apj, 637, 774

\bibitem[{{Cohen} {et~al.}(1999){Cohen}, {Walker}, {Carter}, {Hammersley},
  {Kidger}, \& {Noguchi}}]{cohen99}
{Cohen}, M., {Walker}, R.~G., {Carter}, B., {Hammersley}, P., {Kidger}, M., \&
  {Noguchi}, K. 1999, \aj, 117, 1864

\bibitem[{{Corbel} {et~al.}(2000){Corbel}, {Fender}, {Tzioumis}, {Nowak},
  {McIntyre}, {Durouchoux}, \& {Sood}}]{corbel00}
{Corbel}, S., {Fender}, R.~P., {Tzioumis}, A.~K., {Nowak}, M., {McIntyre}, V.,
  {Durouchoux}, P., \& {Sood}, R. 2000, \aap, 359, 251

\bibitem[{{Corbel} \& {Fender}(2002)}]{corbel02}
{Corbel}, S., \& {Fender}, R.~P. 2002, \apjl, 573, L35

\bibitem[{{Corbel} {et~al.}(2003){Corbel}, {Nowak}, {Fender}, {Tzioumis}, \&
  {Markoff}}]{corbel03}
{Corbel}, S., {Nowak}, M.~A., {Fender}, R.~P., {Tzioumis}, A.~K., \& {Markoff},
  S. 2003, \aap, 400, 1007

\bibitem[{{Draine}(2003)}]{draine03}
{Draine}, B.~T. 2003, \araa, 41, 241

\bibitem[{{Dunn} {et~al.}(2008){Dunn}, {Fender}, {K{\"o}rding}, {Cabanac}, \&
  {Belloni}}]{dunn08}
{Dunn}, R.~J.~H., {Fender}, R.~P., {K{\"o}rding}, E.~G., {Cabanac}, C., \&
  {Belloni}, T. 2008, \mnras, 387, 545

\bibitem[{{Falcke} \& {Biermann}(1995)}]{falcke95}
{Falcke}, H., \& {Biermann}, P.~L. 1995, \aap, 293, 665

\bibitem[{{Fender}(2001)}]{fender01}
{Fender}, R.~P. 2001, \mnras, 322, 31

\bibitem[{{Flaherty} {et~al.}(2007){Flaherty}, {Pipher}, {Megeath}, {Winston},
  {Gutermuth}, {Muzerolle}, {Allen}, \& {Fazio}}]{flaherty07}
{Flaherty}, K.~M., {Pipher}, J.~L., {Megeath}, S.~T., {Winston}, E.~M.,
  {Gutermuth}, R.~A., {Muzerolle}, J., {Allen}, L.~E., \& {Fazio}, G.~G. 2007,
  \apj, 663, 1069

\bibitem[{{Gallo} {et~al.}(2007){Gallo}, {Migliari}, {Markoff}, {Tomsick},
  {Bailyn}, {Berta}, {Fender}, \& {Miller-Jones}}]{gallo07}
{Gallo}, E., {Migliari}, S., {Markoff}, S., {Tomsick}, J.~A., {Bailyn}, C.~D.,
  {Berta}, S., {Fender}, R., \& {Miller-Jones}, J.~C.~A. 2007, \apj, 670, 600

\bibitem[{{Gandhi}(2009)}]{g09_rmsflux}
{Gandhi}, P. 2009, \apjl, 697, L167

\bibitem[{{Gandhi} {et~al.}(2010){Gandhi}, {Dhillon}, {Durant}, {Fabian},
  {Kubota}, {Makishima}, {Malzac}, {Marsh}, {Miller}, {Shahbaz}, {Spruit}, \&
  {Casella}}]{g10}
{Gandhi}, P., {Dhillon}, V.~S., {Durant}, M., {Fabian}, A.~C., {Kubota}, A.,
  {Makishima}, K., {Malzac}, J., {Marsh}, T.~R., {Miller}, J.~M., {Shahbaz},
  T., {Spruit}, H.~C., \& {Casella}, P. 2010, \mnras, 407, 2166 (G10)

\bibitem[{{Gandhi} {et~al.}(2008){Gandhi}, {Makishima}, {Durant}, {Fabian},
  {Dhillon}, {Marsh}, {Miller}, {Shahbaz}, \& {Spruit}}]{g08}
{Gandhi}, P., {Makishima}, K., {Durant}, M., {Fabian}, A.~C., {Dhillon}, V.~S.,
  {Marsh}, T.~R., {Miller}, J.~M., {Shahbaz}, T., \& {Spruit}, H.~C. 2008,
  \mnras, 390, L29

\bibitem[{{Heinz} \& {Sunyaev}(2003)}]{heinz03}
{Heinz}, S., \& {Sunyaev}, R.~A. 2003, \mnras, 343, L59

\bibitem[{{Hynes} {et~al.}(2003){Hynes}, {Steeghs}, {Casares}, {Charles}, \&
  {O'Brien}}]{hynes03_gx339}
{Hynes}, R.~I., {Steeghs}, D., {Casares}, J., {Charles}, P.~A., \& {O'Brien},
  K. 2003, \apjl, 583, L95

\bibitem[{{Jarrett} {et~al.}(2011){Jarrett}, {Cohen}, {Masci}, {Wright},
  {Stern}, {Benford}, {Blain}, {Carey}, {Cutri}, {Eisenhardt}, {Lonsdale},
  {Mainzer}, {Marsh}, {Padgett}, {Petty}, {Ressler}, {Skrutskie}, {Stanford},
  {Surace}, {Tsai}, {Wheelock}, \& {Yan}}]{jarrett11}
{Jarrett}, T.~H., {Cohen}, M., {Masci}, F., {Wright}, E., {Stern}, D.,
  {Benford}, D., {Blain}, A., {Carey}, S., {Cutri}, R.~M., {Eisenhardt}, P.,
  {Lonsdale}, C., {Mainzer}, A., {Marsh}, K., {Padgett}, D., {Petty}, S.,
  {Ressler}, M., {Skrutskie}, M., {Stanford}, S., {Surace}, J., {Tsai}, C.~W.,
  {Wheelock}, S., \& {Yan}, D.~L. 2011, \apj, 735, 112

\bibitem[{{Kellermann} \& {Pauliny-Toth}(1969)}]{kellermann69}
{Kellermann}, K.~I., \& {Pauliny-Toth}, I.~I.~K. 1969, \apjl, 155, L71+

\bibitem[{{K{\"o}rding} {et~al.}(2006){K{\"o}rding}, {Fender}, \&
  {Migliari}}]{koerding06}
{K{\"o}rding}, E.~G., {Fender}, R.~P., \& {Migliari}, S. 2006, \mnras, 369,
  1451

\bibitem[{{Lutz} {et~al.}(1996){Lutz}, {Feuchtgruber}, {Genzel}, {Kunze},
  {Rigopoulou}, {Spoon}, {Wright}, {Egami}, {Katterloher}, {Sturm},
  {Wieprecht}, {Sternberg}, {Moorwood}, \& {de Graauw}}]{lutz96}
{Lutz}, D., {Feuchtgruber}, H., {Genzel}, R., {Kunze}, D., {Rigopoulou}, D.,
  {Spoon}, H.~W.~W., {Wright}, C.~M., {Egami}, E., {Katterloher}, R., {Sturm},
  E., {Wieprecht}, E., {Sternberg}, A., {Moorwood}, A.~F.~M., \& {de Graauw},
  T. 1996, \aap, 315, L269

\bibitem[{{Mainzer} {et~al.}(2011){Mainzer}, {Bauer}, {Grav}, {Masiero},
  {Cutri}, {Dailey}, {Eisenhardt}, {McMillan}, {Wright}, {Walker}, {Jedicke},
  {Spahr}, {Tholen}, {Alles}, {Beck}, {Brandenburg}, {Conrow}, {Evans},
  {Fowler}, {Jarrett}, {Marsh}, {Masci}, {McCallon}, {Wheelock}, {Wittman},
  {Wyatt}, {DeBaun}, {Elliott}, {Elsbury}, {Gautier}, {Gomillion}, {Leisawitz},
  {Maleszewski}, {Micheli}, \& {Wilkins}}]{neowise}
{Mainzer}, A., {Bauer}, J., {Grav}, T., {Masiero}, J., {Cutri}, R.~M.,
  {Dailey}, J., {Eisenhardt}, P., {McMillan}, R.~S., {Wright}, E., {Walker},
  R., {Jedicke}, R., {Spahr}, T., {Tholen}, D., {Alles}, R., {Beck}, R.,
  {Brandenburg}, H., {Conrow}, T., {Evans}, T., {Fowler}, J., {Jarrett}, T.,
  {Marsh}, K., {Masci}, F., {McCallon}, H., {Wheelock}, S., {Wittman}, M.,
  {Wyatt}, P., {DeBaun}, E., {Elliott}, G., {Elsbury}, D., {Gautier}, IV, T.,
  {Gomillion}, S., {Leisawitz}, D., {Maleszewski}, C., {Micheli}, M., \&
  {Wilkins}, A. 2011, \apj, 731, 53

\bibitem[{{Markoff} {et~al.}(2001){Markoff}, {Falcke}, \& {Fender}}]{markoff01}
{Markoff}, S., {Falcke}, H., \& {Fender}, R. 2001, \aap, 372, L25

\bibitem[{{Markoff} {et~al.}(2003){Markoff}, {Nowak}, {Corbel}, {Fender}, \&
  {Falcke}}]{markoff03}
{Markoff}, S., {Nowak}, M., {Corbel}, S., {Fender}, R., \& {Falcke}, H. 2003,
  \aap, 397, 645

\bibitem[{{Meier} {et~al.}(2001){Meier}, {Koide}, \& {Uchida}}]{meier01}
{Meier}, D.~L., {Koide}, S., \& {Uchida}, Y. 2001, Science, 291, 84

\bibitem[{{Migliari} {et~al.}(2007){Migliari}, {Tomsick}, {Markoff}, {Kalemci},
  {Bailyn}, {Buxton}, {Corbel}, {Fender}, \& {Kaaret}}]{migliari07}
{Migliari}, S., {Tomsick}, J.~A., {Markoff}, S., {Kalemci}, E., {Bailyn},
  C.~D., {Buxton}, M., {Corbel}, S., {Fender}, R.~P., \& {Kaaret}, P. 2007,
  \apj, 670, 610

\bibitem[{{Migliari} {et~al.}(2010){Migliari}, {Tomsick}, {Miller-Jones},
  {Heinz}, {Hynes}, {Fender}, {Gallo}, {Jonker}, \& {Maccarone}}]{migliari10}
{Migliari}, S., {Tomsick}, J.~A., {Miller-Jones}, J.~C.~A., {Heinz}, S.,
  {Hynes}, R.~I., {Fender}, R.~P., {Gallo}, E., {Jonker}, P.~G., \&
  {Maccarone}, T.~J. 2010, \apj, 710, 117

\bibitem[{{Motch} {et~al.}(1982){Motch}, {Ilovaisky}, \& {Chevalier}}]{motch82}
{Motch}, C., {Ilovaisky}, S.~A., \& {Chevalier}, C. 1982, \aap, 109, L1

\bibitem[{{Mu{\~n}oz-Darias} {et~al.}(2008){Mu{\~n}oz-Darias}, {Casares}, \&
  {Mart{\'{\i}}nez-Pais}}]{munoz-darias08}
{Mu{\~n}oz-Darias}, T., {Casares}, J., \& {Mart{\'{\i}}nez-Pais}, I.~G. 2008,
  \mnras, 385, 2205

\bibitem[{{Pe'er} \& {Casella}(2009)}]{peercasella09}
{Pe'er}, A., \& {Casella}, P. 2009, \apj, 699, 1919

\bibitem[{{Polko} {et~al.}(2010){Polko}, {Meier}, \& {Markoff}}]{polko10}
{Polko}, P., {Meier}, D.~L., \& {Markoff}, S. 2010, \apj, 723, 1343

\bibitem[{{Rahoui} {et~al.}(2010){Rahoui}, {Chaty}, {Rodriguez}, {Fuchs},
  {Mirabel}, \& {Pooley}}]{rahoui10}
{Rahoui}, F., {Chaty}, S., {Rodriguez}, J., {Fuchs}, Y., {Mirabel}, I.~F., \&
  {Pooley}, G.~G. 2010, \apj, 715, 1191

\bibitem[{{Rahoui} {et~al.}(2011){Rahoui}, {Lee}, {Heinz}, {Hines},
  {Pottschmidt}, {Wilms}, \& {Grinberg}}]{rahoui11}
{Rahoui}, F., {Lee}, J.~C., {Heinz}, S., {Hines}, D.~C., {Pottschmidt}, K.,
  {Wilms}, J., \& {Grinberg}, V. 2011, \apj, 736, 63

\bibitem[{{Russell} {et~al.}(2006){Russell}, {Fender}, {Hynes}, {Brocksopp},
  {Homan}, {Jonker}, \& {Buxton}}]{russell06}
{Russell}, D.~M., {Fender}, R.~P., {Hynes}, R.~I., {Brocksopp}, C., {Homan},
  J., {Jonker}, P.~G., \& {Buxton}, M.~M. 2006, \mnras, 371, 1334

\bibitem[{{Rybicki} \& {Lightman}(1979)}]{rybickilightman}
{Rybicki}, G.~B., \& {Lightman}, A.~P. 1979, {Radiative processes in
  astrophysics} (New York, Wiley-Interscience, 1979.~393 p.)

\bibitem[{{Shahbaz} {et~al.}(2001){Shahbaz}, {Fender}, \&
  {Charles}}]{shahbaz01}
{Shahbaz}, T., {Fender}, R., \& {Charles}, P.~A. 2001, \aap, 376, L17

\bibitem[{{Shidatsu} {et~al.}(2011){Shidatsu}, {Ueda}, {Tazaki}, {Yoshikawa},
  {Nagayama}, {Nagata}, {Oi}, {Yamaoka}, {Takahashi}, {Kubota}, {Cottam},
  {Remillard}, \& {Negoro}}]{shidatsu11}
{Shidatsu}, M., {Ueda}, Y., {Tazaki}, F., {Yoshikawa}, T., {Nagayama}, T.,
  {Nagata}, T., {Oi}, N., {Yamaoka}, K., {Takahashi}, H., {Kubota}, A.,
  {Cottam}, J., {Remillard}, R., \& {Negoro}, H. 2011, PASJ in press, arXiv:1105.3586

\bibitem[{{Titarchuk}(1994)}]{comptt}
{Titarchuk}, L. 1994, \apj, 434, 570

\bibitem[{{Tomsick} {et~al.}(2004){Tomsick}, {Bailyn}, {Buxton}, {Corbel},
  {Fender}, {Markoff}, {Jimenez-Garate}, {Kaaret}, \& {Kalemci}}]{tomsick04}
{Tomsick}, J.~A., {Bailyn}, C.~D., {Buxton}, M.~M., {Corbel}, S., {Fender},
  R.~P., {Markoff}, S., {Jimenez-Garate}, M., {Kaaret}, P., \& {Kalemci}, E.
  2004, in Bulletin of the American Astronomical Society, Vol.~36, American
  Astronomical Society Meeting Abstracts, 104.04--+

\bibitem[{{Trippe} {et~al.}(2011){Trippe}, {Krips}, {Pietu}, {Neri}, {Winters},
  {Gueth}, {Bremer}, {Salome}, {Moreno}, {Boissier}, \& {Fontani}}]{trippe11}
{Trippe}, S., {Krips}, M., {Pietu}, V., {Neri}, R., {Winters}, J.~M., {Gueth},
  F., {Bremer}, M., {Salome}, P., {Moreno}, R., {Boissier}, J., \& {Fontani},
  F. 2011, A\&A in press, arXiv:1107.5456

\bibitem[{{Vaughan} {et~al.}(2003){Vaughan}, {Edelson}, {Warwick}, \&
  {Uttley}}]{vaughan03}
{Vaughan}, S., {Edelson}, R., {Warwick}, R.~S., \& {Uttley}, P. 2003, \mnras,
  345, 1271

\bibitem[{{Vrtilek}(2008)}]{vrtilek08}
{Vrtilek}, S.~D. 2008, in American Institute of Physics Conference Series, Vol.
  1010, A Population Explosion: The Nature \& Evolution of X-ray Binaries in
  Diverse Environments, ed. {R.~M.~Bandyopadhyay, S.~Wachter, D.~Gelino, \&
  C.~R.~Gelino}, 18--22

\bibitem[{{Wachter}(2008)}]{wachter08}
{Wachter}, S. 2008, in American Institute of Physics Conference Series, Vol.
  1010, A Population Explosion: The Nature \& Evolution of X-ray Binaries in
  Diverse Environments, ed. {R.~M.~Bandyopadhyay, S.~Wachter, D.~Gelino, \&
  C.~R.~Gelino}, 210--214

\bibitem[{{Wright} {et~al.}(2010){Wright}, {Eisenhardt}, {Mainzer}, {Ressler},
  {Cutri}, {Jarrett}, {Kirkpatrick}, {Padgett}, {McMillan}, {Skrutskie},
  {Stanford}, {Cohen}, {Walker}, {Mather}, {Leisawitz}, {Gautier}, {McLean},
  {Benford}, {Lonsdale}, {Blain}, {Mendez}, {Irace}, {Duval}, {Liu}, {Royer},
  {Heinrichsen}, {Howard}, {Shannon}, {Kendall}, {Walsh}, {Larsen}, {Cardon},
  {Schick}, {Schwalm}, {Abid}, {Fabinsky}, {Naes}, \& {Tsai}}]{wise}
{Wright}, E.~L., {Eisenhardt}, P.~R.~M., {Mainzer}, A.~K., {Ressler}, M.~E.,
  {Cutri}, R.~M., {Jarrett}, T., {Kirkpatrick}, J.~D., {Padgett}, D.,
  {McMillan}, R.~S., {Skrutskie}, M., {Stanford}, S.~A., {Cohen}, M., {Walker},
  R.~G., {Mather}, J.~C., {Leisawitz}, D., {Gautier}, III, T.~N., {McLean}, I.,
  {Benford}, D., {Lonsdale}, C.~J., {Blain}, A., {Mendez}, B., {Irace}, W.~R.,
  {Duval}, V., {Liu}, F., {Royer}, D., {Heinrichsen}, I., {Howard}, J.,
  {Shannon}, M., {Kendall}, M., {Walsh}, A.~L., {Larsen}, M., {Cardon}, J.~G.,
  {Schick}, S., {Schwalm}, M., {Abid}, M., {Fabinsky}, B., {Naes}, L., \&
  {Tsai}, C.-W. 2010, \aj, 140, 1868 (W10)

\end{thebibliography}
\end{document}